\documentstyle[preprint,eqsecnum,aps,prb]{revtex}
\begin{document}
\title{Control of the colossal magnetoresistance by strain effect in Nd$_{0.5}$Ca$%
_{0.5}$MnO$_{3}$ thin films.}
\author{E.\ Rauwel-Buzin$^{1}$, W. Prellier$^{1,}$\thanks{%
prellier@ismra.fr}, Ch. Simon$^{1}$, S. Mercone$^{1}$, B. Mercey$^{1}$, B.
Raveau$^{1}$, J. Sebek$^{2}$ and J. Hejtmanek$^{2}$}
\address{$^{1}$ Laboratoire CRISMAT, CNRS\ UMR 6508, Bd du\\
Mar\'{e}chal Juin, 14050 Caen Cedex, FRANCE.\\
$^{2}$ Institute of Physics, Na Slovance 2, 182 21 PRAHA 8, Czech Republic.}
\date{\today}
\maketitle

\begin{abstract}
Thin films of Nd$_{0.5}$Ca$_{0.5}$MnO$_{3}$ manganites with colossal
magnetoresistance (CMR) properties have been synthesized by the Pulsed Laser
Deposition technique on (100)-SrTiO$_{3}$. The lattice parameters of these
manganites and correlatively their CMR properties can be controlled by the
substrate temperature $T_{S}$. The maximum CMR effect at 75K, calculated as
the ratio $\rho (H=0T)/\rho (H=7T)$ is 10$^{4}$ for a deposition temperature
of $T_{S}=680%
%TCIMACRO{\UNICODE[m]{0xb0}}%
%BeginExpansion
{{}^\circ}%
%EndExpansion
C$. Structural studies show that the Nd$_{0.5}$Ca$_{0.5}$MnO$_{3}$ film is
single phase, [010]-oriented and has a pseudocubic symmetry of the
perovskite subcell with $a=3.77$\AA\ at room temperature. We suggest that
correlation between lattice parameters, CMR and substrate temperature $T_{S}$
result mainly from substrate-induced strains which can weaken the
charge-ordered state at low temperature.
\end{abstract}

\newpage

Numerous investigations of the manganites Ln$_{1-x}$Ca$_{x}$MnO$_{3}$
(Ln=lanthanides) have been performed last years on bulk ceramics, showing
that the colossal magnetoresistance (CMR) in these materials originates from
a competition between metallic ferromagnetism (FM) and charge-ordered (CO)
insulating antiferromagnetism (AFM) \cite{1,2}. The magnetotransport
properties of these compounds are sensitive to the sample nature- i.e.
single crystal, ceramic, thin film; the studies of manganite thin films \cite
{3,4,5,6,7,8,9} and their properties being often different from the bulk
ceramics or single crystals.

The great potentiality of thin films for device applications makes that
their magnetotransport properties need to be investigated in detail in the
future. In this respect, our recent studies of Pr$_{0.5}$Ca$_{0.5}$MnO$_{3}$
thin films \cite{8,9} have shown that the substrate-induced strains play a
key role in the appearance of CMR by decreasing the stability of the
charge-ordered (CO) state. Keeping in mind the key effect of the A-site
cation size observed for bulk manganites\cite{10}, it is of a great interest
to extend previous studies to Nd$_{0.5}$Ca$_{0.5}$MnO$_{3}$ (NCMO) thin
films despite that unreasonably large magnetic field (%
%TCIMACRO{\TEXTsymbol{>}}%
%BeginExpansion
\mbox{$>$}%
%EndExpansion
20 T) were reported to melt the AFM charge ordered state below 200 K in bulk
samples\cite{11}. Moreover no direct confirmation of a complete conversion
to FM metallic state was claimed.

In the present letter, we show that Nd$_{0.5}$Ca$_{0.5}$MnO$_{3}$ thin
films, grown on SrTiO$_{3}$ by pulsed laser deposition (PLD), exhibit
spectacular CMR effect at 7T, involving an insulator-to-metal transition
around 100K. We demonstrate that the perovskite subcell at room temperature
(RT) varies with the temperature of the substrate during deposition, and
correlatively, the CMR effect is controlled by the growth conditions. The
correlations between CMR and lattice parameters are interpreted in terms of
strains effects due to the substrate and oxygen non-stoechiometry.

Dense targets of Nd$_{0.5}$Ca$_{0.5}$MnO$_{3}$ were synthesized via the
standard ceramic methods. The PLD apparatus has been described in detail
elsewhere\cite{8,9}. Briefly, an excimer laser (Lambda Physics, with a $%
\lambda =248nm$ pulse rate of $3Hz$, was focussed onto a rotating target.
1200\AA\ thick films were deposited on (100)-SrTiO$_{3}$ (STO) substrates
heated during deposition at constant temperature ranging from 590%
%TCIMACRO{\UNICODE{0xb0}}%
%BeginExpansion
${{}^\circ}$%
%EndExpansion
C to 770%
%TCIMACRO{\UNICODE{0xb0}}%
%BeginExpansion
${{}^\circ}$%
%EndExpansion
C under an oxygen pressure of about 300mTorr. Details of the growth
parameters will be discussed hereafter. Structural characterizations were
done using an X-ray diffraction (XRD) facility (Seifert 3000 P for $\Theta $%
-2$\Theta $ and Philips X'Pert MRD for in-plane measurements, Cu K$\alpha $
with $\lambda =1.5406$\AA ). Resistivity ($\rho $) of NCMO films was
measured with a PPMS Quantum Design as a function of magnetic field ($H$)
and temperature ($T$) using 4-probe contact. The Energy Dispersive
Spectroscopy analyses have shown that the cationic composition the film is
homogeneous and remains very close to the nominal one ''Nd$_{0.5\pm 0.02}$Ca$%
_{0.5\pm 0.02}$MnO$_{x}$'' within experimental error.

In order to obtain simultaneously as good crystallization and full oxidized
films, we have performed the synthesis under 400mtorr of O$_{2}$ since the
preparation of manganites with high Mn valency usually require a high oxygen
pressure, or a post-annealed treatment\cite{12}. The best results concerning
CMR were obtained for a substrate temperature ($T_{S}$) of $680%
%TCIMACRO{\UNICODE[m]{0xb0}}%
%BeginExpansion
{{}^\circ}%
%EndExpansion
C$ and, consequently, following study was focused on this sample. The
corresponding XRD patterns confirm a single phase (Fig. 1) with an
out-of-plane parameter multiple of $a_{C}$ (where $a_{C}$ refers to the
ideal cubic parameter 3.9\AA ). The high intensity of the peak indicating a
good crystallization for the film is independently confirmed by the
full-width-at half maximum of the rocking-curve equal to 0.25%
%TCIMACRO{\UNICODE{0xb0} }%
%BeginExpansion
${{}^\circ}$%
%EndExpansion
(not shown). The sharp peaks 90%
%TCIMACRO{\UNICODE{0xb0}}%
%BeginExpansion
${{}^\circ}$%
%EndExpansion
-separated in the $\Phi $-scan of the NCMO film (see inset of Fig. 1) make
evident the existence of a complete in-plane orientation of the film. In a
sake of clarity the diffraction peak around $2\Theta =47%
%TCIMACRO{\UNICODE[m]{0xb0}}%
%BeginExpansion
{{}^\circ}%
%EndExpansion
$ (see Fig.1) has been indexed in the perovskite cell as a $(002)_{C}$
(where c refers to the cubic perovskite cell). In fact, the electron
diffraction study of selected NCMO films confirmed that they are single
phase. We have found that they are [010]-oriented referring to the $Pnma$
space group\cite{13} as previously reported for Pr$_{0.5}$Ca$_{0.5}$MnO$_{3}$
thin films grown on (100)-SrTiO$_{3}$ substrates\cite{8}. The details of the
transmission electron microscopy study are currently undertaken and will be
published elsewhere. The in-plane parameter has been evaluated by
considering the $(103)_{C}$ reflection; It is remarkable that the parameters
of the perovskite subcell are very close to each other, i.e. $3.775$\AA\ and 
$3.765\AA $\ for the in-plane and out of plane parameters respectively. This
indicates a pseudo cubic symmetry which is similar to the bulk Nd$_{0.5}$Ca$%
_{0.5}$MnO$_{3}$, but the latter exhibits significantly larger cell
parameters at RT ($a/\sqrt{2}=3.821$\AA , $b/2=3.797$\AA\ and $c/\sqrt{2}%
=3.805$\AA \cite{13}). But as more important we emphasize the substantial
variation of RT lattice parameters with the deposition temperature TS, in
the range 620%
%TCIMACRO{\UNICODE{0xb0}}%
%BeginExpansion
${{}^\circ}$%
%EndExpansion
C to 780%
%TCIMACRO{\UNICODE{0xb0}}%
%BeginExpansion
${{}^\circ}$%
%EndExpansion
C. This dependence is shown in Figure 2a. Indeed for $T_{S}$ ranging from 620%
%TCIMACRO{\UNICODE{0xb0} }%
%BeginExpansion
${{}^\circ}$%
%EndExpansion
to 680%
%TCIMACRO{\UNICODE{0xb0}}%
%BeginExpansion
${{}^\circ}$%
%EndExpansion
C the symmetry remains pseudocubic and the cell parameters decrease almost
linearly with increasing $T_{S}$ . Contrary to that for $T_{S}>680%
%TCIMACRO{\UNICODE[m]{0xb0}}%
%BeginExpansion
{{}^\circ}%
%EndExpansion
C$, a sudden distortion of the perovskite subcell appears, the in plane and
out of plane parameters being very different, and further increase with $%
T_{S}$. From this observations we deduce that the strains induced by the
substrate are not the only factor which governs the crystallographic nature
of the film and we presume that the oxygen stoechiometry of NCMO thin film
varies with the deposition temperature. This general statement is reinforced
by the evolution of the cell volume versus $T_{S}$ (Fig. 2a) when below 680%
%TCIMACRO{\UNICODE{0xb0}}%
%BeginExpansion
${{}^\circ}$%
%EndExpansion
C the volume decrease likely suggests an oxygen uptake with increasing $T_{S}
$ due to increasing Mn valency (Mn$^{4+}$ for Mn$^{3+}$ substitution),
whereas above 680%
%TCIMACRO{\UNICODE{0xb0}}%
%BeginExpansion
${{}^\circ}$%
%EndExpansion
C the volume increase can be explained by an oxygen departure -i.e. Mn$^{3+}$
for Mn$^{4+}$ substitution.

Keeping in mind that the substrate-induced strains may crucially affect the
temperature and stability of phase transitions , we may expect that the CMR
properties of films will depend upon the crystallographic nature of the room
temperature form and, consequently, upon the deposition temperature $T_{S}$.
We have focussed our study on the best crystallized films deposited under
400mtorr of O$_{2}$ in the temperature range of 620%
%TCIMACRO{\UNICODE{0xb0}}%
%BeginExpansion
${{}^\circ}$%
%EndExpansion
C and 680%
%TCIMACRO{\UNICODE{0xb0}}%
%BeginExpansion
${{}^\circ}$%
%EndExpansion
C (below 620%
%TCIMACRO{\UNICODE{0xb0}}%
%BeginExpansion
${{}^\circ}$%
%EndExpansion
C, the crystallization of the film is poor). As stated above, the best
result was observed for the film grown at 680%
%TCIMACRO{\UNICODE{0xb0}}%
%BeginExpansion
${{}^\circ}$%
%EndExpansion
C. The $\rho (T)$ curve registered in absence of magnetic field and at 7T
are shown in Fig.3a first order insulator to metal transitions,
characterized by a hysteretic resistivity jump of several orders of
magnitude (see also $\rho (H)$ in the inset Fig.3a). According to our
knowledge such a sharp transition has never been detected before in thin
films. In context to the metallic ground state at low temperatures,  let us
underline the very low value of residual resistivity ($\symbol{126}30m\Omega
cm$) which independently confirms both the high quality of the film and the
complete transformation to FM metallic phase.

Thus based on resistivity measurements we deduce a phase diagram- see
Fig.3b. Using the hysteresis cycles in the field ramping up ($H_{C}^{+}$)
and ramping down ($H_{C}^{-}$), we define the thermodynamic filed ($H_{C}$)
as the average of ($H_{C}^{+}$) and ($H_{C}^{-}$). In Fig. 3b, $H_{C}$ shows
a monotonous increase with temperature when T%
%TCIMACRO{\TEXTsymbol{<}}%
%BeginExpansion
\mbox{$<$}%
%EndExpansion
150K. $H_{C}$ is estimated to be 3T at 25 K, which is 6-7 times lower than
the 25T found in the bulk material with the same chemical composition \cite
{11}. In fact, our phase diagram of 680%
%TCIMACRO{\UNICODE{0xb0}}%
%BeginExpansion
${{}^\circ}$%
%EndExpansion
C deposited NCMO film is very different from the bulk where the magnetic
field required to melt the charge ordered state is much higher at each
temperature\cite{11}.

The second important point of the paper concerns the influence of the
deposition temperature $T_{S}$ of NCMO films up on the CMR effect. The
evolution of the resistance ratio $RR=\rho (H=0T)/\rho (H=7T)$ vs. $T_{S}$
(Fig.2b) shows that the $RR$ increases as $T_{S}$ increases up to $T_{S}=680%
%TCIMACRO{\UNICODE[m]{0xb0}}%
%BeginExpansion
{{}^\circ}%
%EndExpansion
C$ and suddenly decreases for $T_{S}>680%
%TCIMACRO{\UNICODE[m]{0xb0}}%
%BeginExpansion
{{}^\circ}%
%EndExpansion
C$. It is straightforward to correlate this evolution to the lattice
parameters :

-the CMR effect increases in the pseudocubic region as the lattice
parameters at room temperature decrease reaching a maximum value of 10$^{4}$
for a minimum average lattice parameter of 3.77\AA\ (corresponding to $%
T_{S}=680%
%TCIMACRO{\UNICODE[m]{0xb0}}%
%BeginExpansion
{{}^\circ}%
%EndExpansion
C$)

-then it drops abruptly in the distorted lattice region as the parameters
increase. Similar evolutions are observed for $RR$ measured at $T=100K$ and $%
125K$, while the maximum $RR$ value rapidly decreases, as one approaches $%
T_{CO}$ ($T_{CO}$=250 K).

The dramatic influence of the substrate temperature T$_{S}$ upon both,
lattice parameters and CMR properties of $Nd_{0.5}Ca_{0.5}MnO_{3}$ thin
films suggests that two factors influence their properties , the oxygen
stoechiometry and the strains induced by the substrate. Concerning the
influence of the oxygen stoechiometry, the study on bulk ceramics by
Frontera et al. \cite{14} has shown, that an oxygen overstoechiometry in $%
Nd_{0.5}Ca_{0.5}MnO_{3+\delta }$ leads to a significant variation of the
structural and magnetic properties, namely the formally excess oxygen d
induces the increase of ferromagnetic interactions at low temperatures.
Nevertheless, the lattice and CMR effects encountered for our NCMO films are
huge compared to the $Nd_{0.5}Ca_{0.5}MnO_{3+\delta }$ ceramics, so we do
not suppose that oxygen non-stoechiometry is the only factor controlling the
properties of our films. The second factor dealing with the strains induced
by the substrate, previously observed for other CO manganites \cite{8,9,15},
has at least the same prominence since, as previously observed, the
substrate-layer strain is susceptible e.g. weakening the CO state, by
blocking the variation of the in-plane lattice parameters with increasing
temperature\cite{9}, thus favoring the CMR effect.

To conclude, high quality Nd$_{0.5}$Ca$_{0.5}$MnO$_{3}$ thin films, with
exceptional CMR properties have been grown using pulsed laser deposition.
The room temperature lattice parameters and their CMR properties of films
are correlated and can be controlled by the deposition temperature $T_{S}$.
We suggest that observed correlations between structural and transport
properties in NCMO films result predominantly from substrate-induced strains
which can destabilize the charge-ordered state.

Acknowledgements: W. Prellier thanks Dr C. Frontera for helpful discussions.

\newpage

Figures Captions:

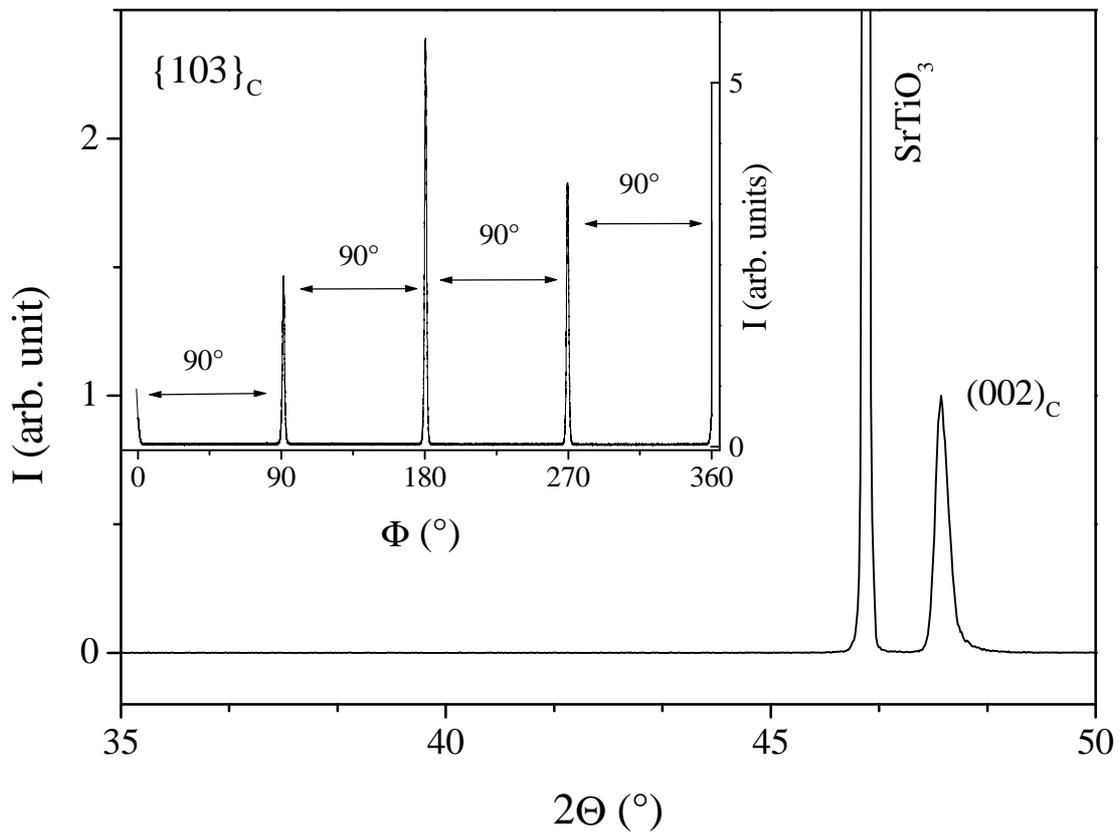
Fig.1: Room temperature $\Theta $-2$\Theta $ XRD pattern in the range 35-50%
%TCIMACRO{\UNICODE{0xb0} }%
%BeginExpansion
${{}^\circ}$%
%EndExpansion
of NCMO/STO. The inset depicts the $\Phi $-scan recorded around the \{103\}$%
_{C}$.

Fig.2a: Evolution of the lattice parameters and the volume of the cell vs.
deposition temperature for films grown under 400mTorr of O$_{2}$.

Fig.2b: CMR effect calculated as $\rho (H=0)/\rho (H=7T)$ vs. the deposition
temperature (400mTorrs of O$_{2}$ was used).

Fig.3a: $\rho $(T) under different applied magnetic fields for an optimized
film ($T_{S}=680%
%TCIMACRO{\UNICODE[m]{0xb0}}%
%BeginExpansion
{{}^\circ}%
%EndExpansion
C$). The inset shows the $\rho $(H) at 75K. Note the sharp transition and
the CMR effect of 11 orders of magnitude at 7T. Field is applied parallel to
the substrate plane. The arrows indicate the increase and the decrease in T.

Fig.3b: Corresponding $T-H$ phase diagram. The dashed area indicates the
hysteresis region where CO and FM phases coexist.

\end{document}